\documentclass[12pt]{article}

\usepackage{scicite}
\usepackage{setspace}

\usepackage{times}
\usepackage{amsmath,amssymb}
\usepackage{graphicx}

\newenvironment{sciabstract}{%
\begin{quote} \bf}
{\end{quote}}

\title{Observation of Pauli blocking in light scattering from quantum degenerate fermions } 

\author
{Amita.~B.~Deb,$^{1\ast}$  Niels.~Kj{\ae}rgaard$^{1}$\\
\normalsize{$^{1}$Dodd-Walls Centre and the Department of Physics, University of Otago, New Zealand.}\\
\normalsize{$^\ast$To whom correspondence should be addressed; E-mail:  amita.deb@otago.ac.nz.}
}

\date{}

\begin{document}

\baselineskip24pt


\maketitle 


\begin{sciabstract}
\singlespacing
The Pauli exclusion principle forbids indistinguishable fermions to occupy the same quantum mechanical state. Its implications are profound and it for example accounts for the electronic shell structure of atoms. Here we perform measurements on the scattering of off-resonant light from ultracold gasses of fermionic atoms. For Fermi gases in the quantum degenerate regime, we observe a marked suppression in light scattering as compared to a similarly prepared thermal Bose gas. We attribute the observed increased transmission of light through the quantum degenerate Fermi gas to Pauli blocking, where Fermi-Dirac statistics causes atoms to occupy a large region of the momentum space limiting the number of accessible states for the scattered atom. Our work confirms a longstanding fundamental result in the theory of the optical response of quantum gases and is an important step towards novel cooling and thermometry mechanisms for degenerate Fermi gases.

\end{sciabstract}

\singlespacing

 Interaction of light with ultracold atoms is of paramount importance in emerging quantum technologies based on quantum gases. Understanding collective and quantum effects in the interplay between light with atomic dipoles is a highly active field in atomic physics \cite{Bettles2016,Pellegrino2014,Bromley2016,Rui2020a} with its origin in a seminal paper by Dicke \cite{Dicke1954} which predicted a modification of radiative lifetime of atoms placed within a volume smaller than the cube of the wavelength of light. Another prominent example is the Purcell effect \cite{Purcell1946} where the spontaneous emission rate is modified due to a change in the number of accessible states for photons. This, for example, happens for atoms in optical resonators and lies at the heart of cavity quantum electrodynamics.

\begin{figure}
\begin{center}
         \includegraphics[width=0.8\textwidth]{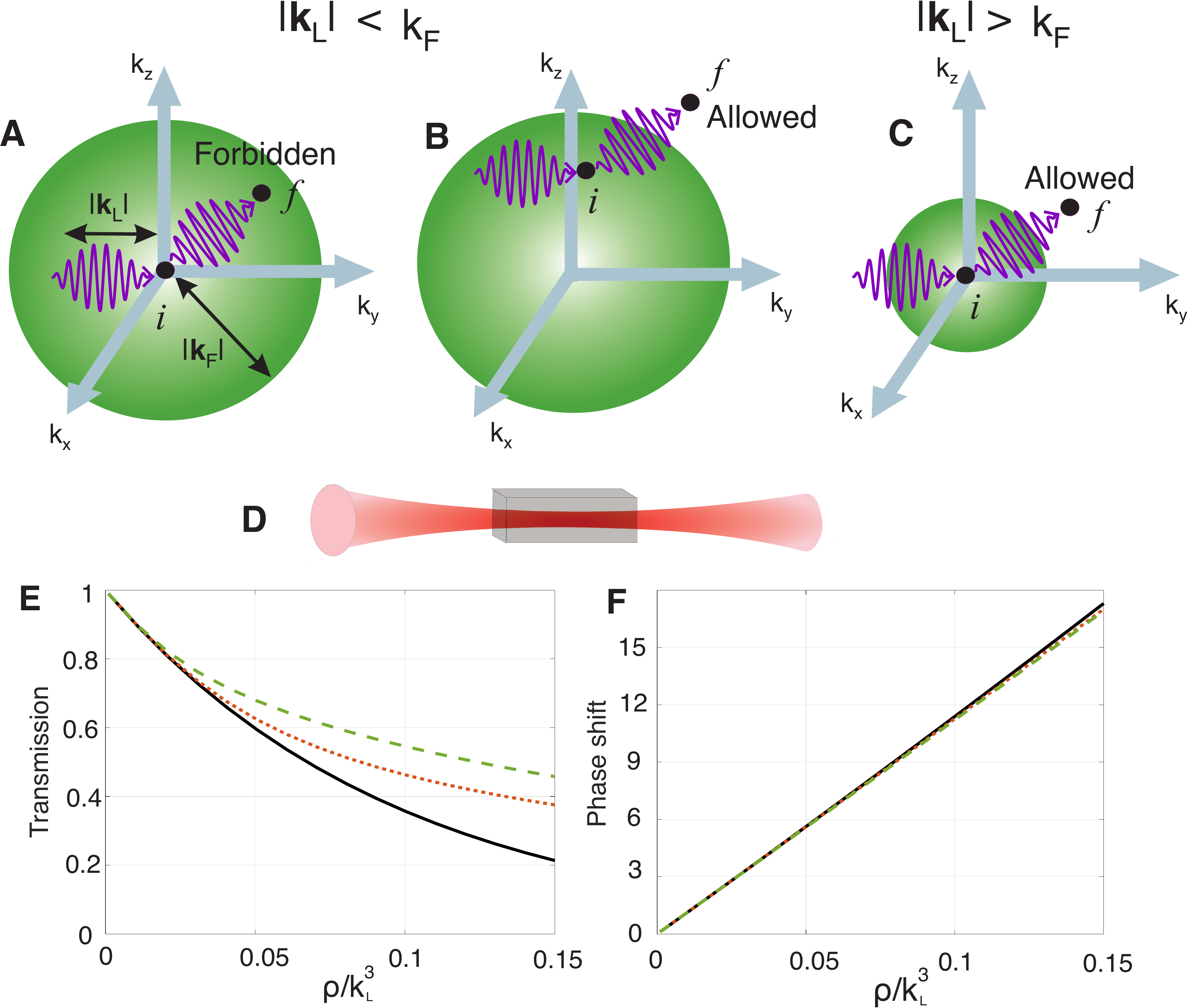}
        \caption{\textbf{(A-C)} show the Fermi spheres at $T = 0$ for a homogeneous Fermi gas, where all states within a sphere of radius $k_F$ are occupied. For  $|\textbf{k}_L| < k_F $, an atom near the centre of the sphere can not scatter an incoming photon, as its final momentum state $f$ is occupied, shown in  \textbf{A}. For an atom near the surface of the sphere, shown in  \textbf{(B)}, photon scattering is allowed. When $|\textbf{k}_L| > k_F$, as in  \textbf{(C)}, photon scattering is more likely to be allowed. \textbf{(D)} shows an off-resonant light beam passing through a homogeneous Fermi gas of finite length. \textbf{(E)} and \textbf{(F)} show the transmission and phase shift of the beam as function of density for T~=~2~$\mu$K (solid line), 400\,nK (dotted line) and 200\,nK (dashed line), where we assumed $^{40}$K atoms, $\Delta = -25\gamma$ and the length of the gas to be $30\,\mu$m. }
        \label{Fig1}
        \end{center}
\end{figure}

In addition to being responsible for the electronic structure of atoms, Pauli blocking of fermions reveals itself as anti-bunching of density-density correlations\cite{Oliver1999,Henny1999} and suppressed density fluctuations \cite{Mueller2010,Sanner2010,Omran2015}. It also prevents colliding fermions from scattering sideways \cite{Thomas2016}. In the context of light-matter interaction, quantum statistical effects in the spontaneous emission rate $\gamma$ in degenerate quantum gases was first considered in \cite{Helmerson1990a} three decades ago. In particular, it was predicted that $\gamma$ is suppressed in a deeply degenerate Fermi gas due to Pauli blocking. As pointed out in \cite{Busch1998,Shuve2009}, a scenario that is more tractable experimentally is the case of light scattering from trapped fermions (see Fig.\ref{Fig1}). The theory of the coherent optical response of a dilute quantum degenerate Bose gas was derived in \cite{Morice1995} whereas a derivation for a low density degenerate Fermi gas was given in \cite{Ruostekoski1999}. The possibility of suppression of light scattering has received continuous interest since the first observation of degenerate Fermi gases \cite{DeMarco1998,DeMarco1999,Sanner2010,Ruostekoski2009,Shuve2009}, but its direct detection in an experiment has remained elusive. 

Here we report on an experimental observation of the suppression of off-resonant light scattering from a degenerate Fermi gas of $^{40}$K. We present measurements of phase shift and transmission of a light beam propagating through a Fermi gas, thereby mapping the optical susceptibility of the medium. Measurements on a thermal bosonic gas  of $^{87}$Rb described by classical Maxwell-Boltzmann distribution under very similar experimental conditions are also reported. We observe a significant increase in transmission for the Fermi gas compared to the thermal Bose gas,  which we attribute to the role of Fermi-Dirac statistics leading to Pauli blocking of light scattering. 

The physical mechanism behind the Pauli blocking of light scattering is schematically represented in Fig. 1A-C where we show the momentum space of a homogeneous Fermi gas at zero temperature for a given point in space. All the points within a radius equal to the Fermi wavevector $|k_F| = (6\pi^2\rho)^{1/3} $ are occupied, where $\rho$ is the atomic density. The change of wave number of an atom upon scattering a photon is $|\Delta k_L| = 2 k_L \sin(\theta/2)$, where $k_L = 2\pi/\lambda$ with $\lambda$ being the wavelength of light and $\theta$ is the angle between the incident direction and the direction of the scattered photon. We see that for $k_F > k_L$, recoil events leading to atoms populating an occupied state of the Fermi sea would be forbidden by the Pauli exclusion principle. To observe this effect clearly, one requires a Fermi energy $E_F$ that is much greater than the photon recoil energy $E_R = \hbar^2k_L^2/2m$, a regime that is experimentally accessible with relatively heavy fermionic atoms like $^{40}$K, as in our case. 

The absorption and phase shift of a light beam propagating through a medium is given by its optical susceptibility, which for a homogeneous  low-density Fermi-Dirac gas made up of two-level atoms was derived in \cite{Ruostekoski1999} to be
  \begin{equation}\label{Fermi_susc}
\chi = \frac{\alpha\rho}{1 - \alpha\rho/3 + \mathcal{C}}.
\end{equation}
Here $\alpha = \frac{-6\pi}{k_L^3[(\delta/\gamma) + i]}$ is the polarizability of the atom, $\gamma = \frac{d^2k^3}{6\pi\epsilon_0\hbar}$ where $d$ is the transition dipole moment, and $\rho$ is the atomic density. The second term in the denominator is the well-known Lorentz-Lorenz local field correction. The term $\mathcal{C}$ arises from spatial correlations between atomic dipoles due to Fermi-Dirac statistics, i.e., the fermionic pair correlation function displaying the well-known anti-bunching behaviour at distances shorter than $1/k_F$ \cite{Mueller2010,Sanner2010,Ruostekoski1999}. Figure 1E and  F show the transmission and phase shift as a function of atomic density for an off-resonant ($\Delta = -25\gamma$) light beam propagating through a homogeneous Fermi gas of length $30\,\mu$m. For a homogeneous Fermi gas with sufficiently high density and at low temperatures, there is a significant suppression of absorption of light compared to a high temperature (classical) gas. There is an accompanying, small radiative shift of the resonance, which bears a very small effect on the phase shift of off-resonant light passing through the medium.

\begin{figure}
\begin{center}
        \includegraphics[width=0.5\textwidth]{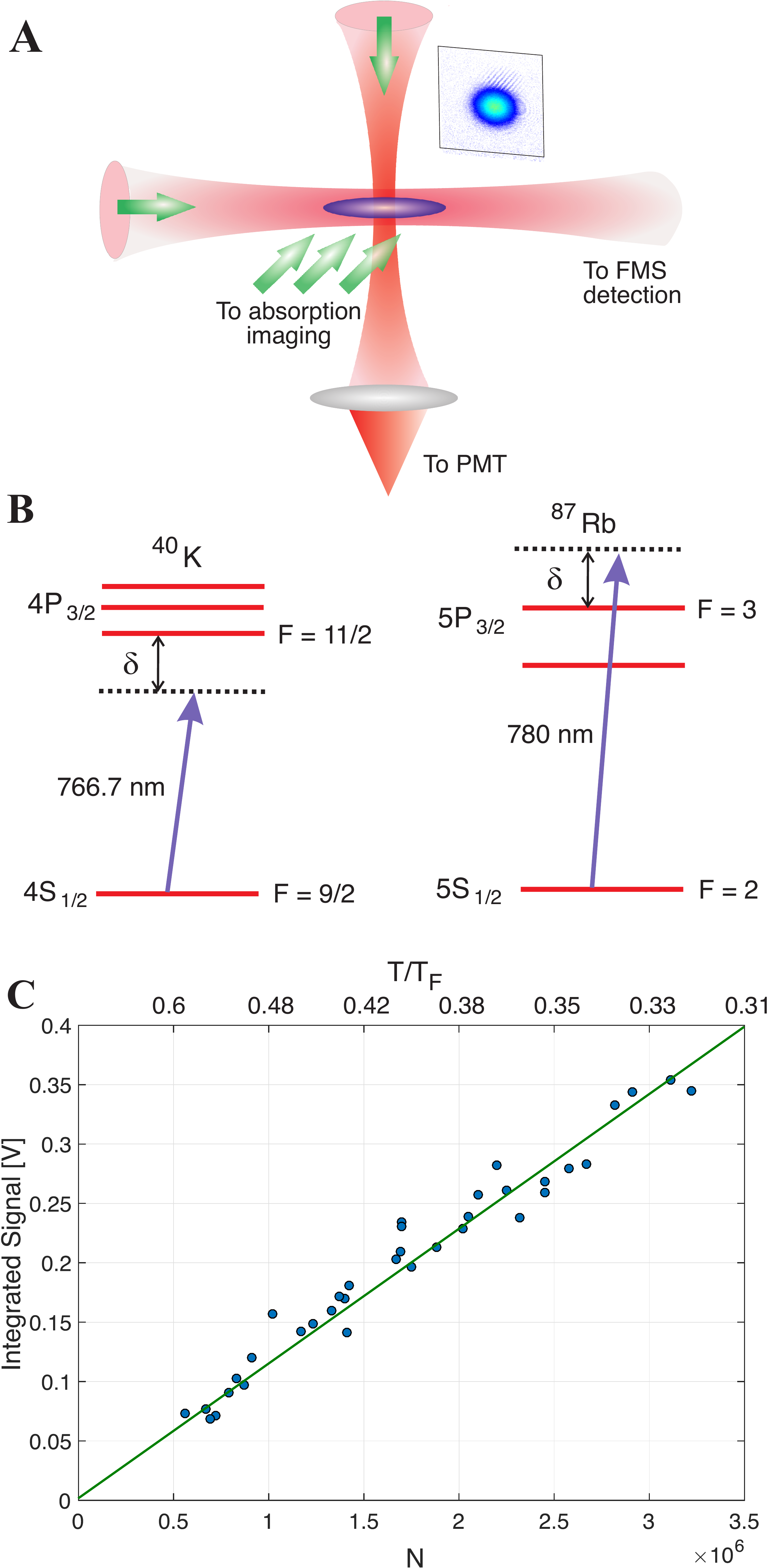}
 \caption{\textbf{(A)} The experimental setup for measuring the dispersive and absorptive response of the gas (see text). Following either of these measurements, the atom number and temperature of the gas are obtained through absorption imaging of in time-of-flight. \textbf{(B)} Atomic levels involved in optical probing of $^{40}$K (left) and $^{87}$Rb (right).  \textbf{(C)} Dispersive response of a Fermi gas as function of atom number with a probe detuning $\delta = -420\,\gamma$. The vertical axis is the FMS signal (see text) that measures the elastically scattered field in the forward direction. The top x-axis shows the scaled temperatures $T/T_F$ (note that the axis is nonlinear). }
        \label{Fig2}
        \end{center}     
\end{figure}

In our experiment, we produce quantum degenerate Fermi gases of $^{40}$K in the ground hyperfine state $|4S_{1/2}; F = 9/2, m_F = 9/2\rangle$ in a harmomic trap provided by an Ioffe-Pritchard (IP) type magnetic trap with radial and axial trapping frequencies of $2\pi \times 260$\,Hz and $2\pi \times 30$\,Hz, respectively. Using a sympathetic cooling procedure with $^{87}$Rb atoms as the coolant, we can produce characteristic $^{40}$K samples with $2.2 \times 10^6$ atoms at temperatures $\sim 420$\,nK, corresponding to $T/T_F \lesssim 0.3$. To map the dispersive response of the atoms, we use the technique of frequency modulation spectroscopy (FMS) \cite{Bjorklund1980, Deb2013}.  Here an electro-optically generated phase-coherent frequency triplet (a carrier and two sidebands) is used as a probe, where only one of the sidebands (the probe sideband) is close to an atomic resonance and is thus perturbed by the atoms. The probe light beam containing the frequency triplet propagates along the axial direction of the atomic cloud (see Figure 2A) and is focussed to a $1/e^2$ waist size of $(25\pm 2)\,\mu$m. We use $\sigma^{+}$ polarized 766.7\,nm light for addressing the cycling transition $|4S_{1/2}; F = 9/2, m_F = 9/2\rangle \rightarrow |4P_{3/2}; F = 11/2, m_F = 11/2\rangle$ of $^{40}$K (see Figure 2b). The FMS signal is sensitive to both extinction and phase shift of the probe sideband. In order to isolate the dispersive response of the atoms, we frequency-detune the probe sideband by $\delta = -420\, \gamma$, such that the optical depth of the sample remains below 0.02 for our highest atomic densities.  The probe sideband has a power of $\sim 700$\,nW and the beam is pulsed for a duration of $4\,\mu$s, leading to $\sim$ 0.01 photon scattering per atom during the probe pulse. Immediately after the probe pulse, the atoms are released from the IP trap and an absorption image is acquired in the transverse direction following a time of flight of 10-15\,ms. From this, we deduce the atom number and the temperature of the sample. The atom number and the temperature of the samples can be independently controlled by adjusting the initial loading of the trap and the final microwave frequency for forced evaporation of the coolant atoms. Figure 2C and D shows the FMS signal as a function of atom number for $^{40}$K samples at 530\,nK. The linear variation of the signal with respect to the atom number indicates that the dispersive response of an atomic medium is practically unmodified by Fermi degeneracy, similar to the homogeneous optical susceptibility model provided by Equation~1.

To measure the imaginary part of the optical susceptibility with a high signal to noise ratio, but low multiple scattering of photons, an optical density in the order of one is optimal \cite{Shuve2009}. This requires us to operate much closer to the atomic resonance than above. We use a single frequency probe beam that propagates along the short axis of the cloud and is focussed to a $1/e^2$ waist size of $(19\pm 2)\,\mu$m. The probe light is polarized perpendicular to the long axis of the cloud and we used a frequency detuning of $\delta = -9\, \gamma$. The light after propagation through the atoms is captured by an optical system with an effective numerical aperture of $\sim 0.095$ and is detected using a photomultiplier tube (PMT) with a small aperture. In this manner we are able to preferentially sample the central part of the atomic distribution, which we verified by measuring optical depths of thermal atomic ensembles with a known in-trap density distribution. The probe beam has a power of $\sim 1$\,nW and is pulsed for a duration of $90\,\mu$s. Under these conditions, an independent atom would scatter around 0.1 photon during the probe pulse. As a result, the perturbation caused by the probe pulse on the temperature and the Fermi sea occupation of the ensemble can be considered negligible, which we have verified using absorption imaging.

\begin{figure}
\begin{center}
        \includegraphics[width=0.60\textwidth]{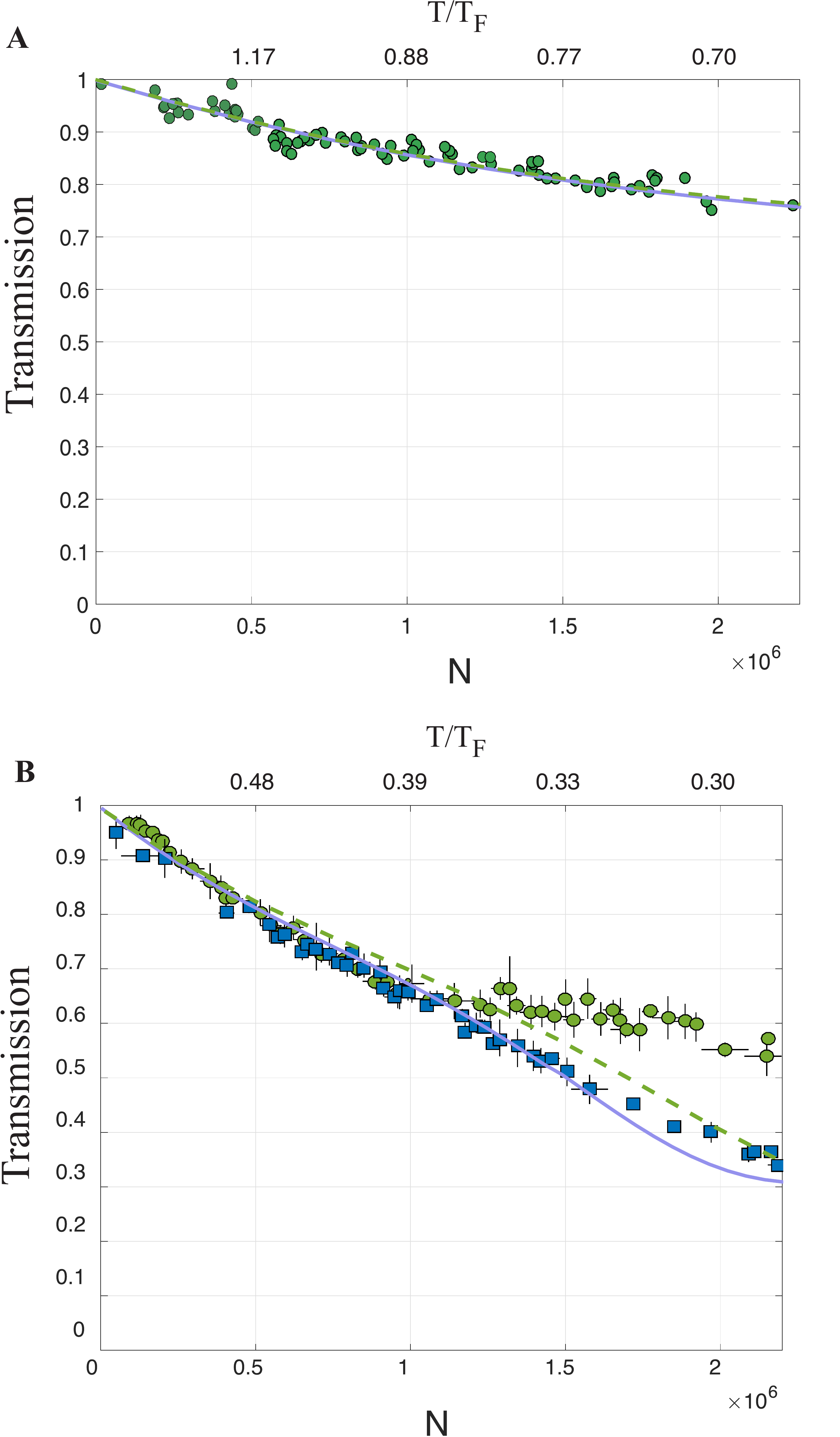}
     
        \caption{\textbf{(A)} Transmitted fraction of light passing through a Fermi gas at temperature 980\,nK as a function of atom number. The top x-axis (nonlinear) shows the scaled temperatures for given atom numbers. \textbf{(B)} Transmitted fraction of light passing through a Fermi gas at temperature 440\,nK (green circles) and a thermal Bose gas at temperature 480\,nK (blue squares) as a function of atom number. Each data point is an average of five experimental realisations. In \textbf{(A)} and \textbf{(B)} , the solid lines are results of the paraxial wave equation model assuming a thermal density distribution and dashed lines are that assuming a polylogarithmic density distribution appropriate for a degenerate Fermi gas. }
        \label{Fig3}
        \end{center}
\end{figure}

We determine the ratio of the light transmitted through $^{40}$K samples measured on the PMT with and without the atoms. Figure 3A shows this ratio as a function of atom number for a  980~($\pm$ 50)\,nK sample. The exponentially decaying transmission profile is well described by a Beer-Lambert picture of extinction of light. In contrast, Fig. 3B presents the results of measurements on samples with a temperature of 440 ($\pm$ 60)\,nK over the same range of atom numbers,  where we observe a clear departure from exponentially decaying behaviour as the number of atoms increases. A crucial difference between the scenarios in Fig. 3A and  B is the degree of quantum degeneracy, as given by $T/T_F$. While Fig. 3A ranges between $T/T_F = 1.2 - 0.7$, Fig. 3B explores a more deeply degenerate regime with $T/T_F = 0.5 - 0.3$. In Fig. 3B, we also show measurements with thermal samples of bosonic $^{87}$Rb atoms at temperature 480 ($\pm$ 50)\,nK, under similar probing conditions.  Compared to the thermal bosonic case, the transmission of light through fermionic atomic clouds is substantially enhanced for higher atom numbers, where the degeneracy is significant. 

Transmission of off-resonant light through a medium of significant density inhomogeneity, as in our case, must be interpreted with care. For off-resonant light, the real part of the optical susceptibility of the atomic medium, and therefore its refractive index, is non-vanishing. A gaussian density distribution of atoms then acts as a gradient index lens that deviates light out of the incident mode.  Such lensing effects can potentially affect transmission measurements for a finite aperture. In order to account for such effects, we solve a paraxial wave equation that models the propagation of light through an inhomogeneous medium \cite{Deb2020}. The solid lines in Fig. 3A and B show the results of simulations where we assumed the gaussian density profile for the medium and ignored the Fermi correlation term $\mathcal{C}$ in the optical susceptibility (Equation 1). The model is in good agreement with our experimental observations both for Fermi gases above $T/T_F \gtrsim 0.5$ and for the thermal bosonic gas. The dashed lines show the results for simulations using the polylogarithmic density profile appropriate for Fermi gases \cite{Ketterle}. Because of the flattening of the fermionic density distribution, our modelling predicts an enhanced transmission relative to a gaussian density distribution. Compared to the experimentally observed enhancement of transmission for degenerate fermions, this is, however, a minor effect. 

The full theoretical modelling of light propagation in a trapped, inhomogeneous, dense and degenerate fermionic medium is a challenging theoretical problem that we have not undertaken in this work. This leaves us unable to compare our observations with numerical predictions. However, we can directly contrast light scattering from a degenerate Fermi gas with a corresponding thermal bosonic gas which a similar classical optical response and for which we can provide adequate modelling. Because we observe a compelling difference between the two cases and since the key attribute differentiating the two cases is their quantum statistics, our observations provide a strong signature of of Pauli blocking of light scattering in a Fermi gas.

In summary, we presented an experimental observation of the suppression of the absorptive optical response of degenerate Fermi gases, confirming a three decade old prediction. In contrast, the dispersive response was observed to be unaltered by Fermi degeneracy. These observations are qualitatively consistent with the optical response derived in \cite{Ruostekoski1999} for a homogeneous Fermi-Dirac gas. Our observations provide new insights into the fundamental theoretical problem concerning the optical response of quantum gases, and paves the way for novel cooling and thermometry schemes for fermions deep below the Fermi temperature, and quantun non-demolition measurements with reduced dissipation.


\begin{thebibliography}{10}
	
	\bibitem{Bettles2016}
	R.~J. Bettles, S.~A. Gardiner, C.~S. Adams, {\it Phys. Rev. Lett.\/} {\bf 116},
	103602 (2016).
	
	\bibitem{Pellegrino2014}
	J.~Pellegrino, {\it et~al.\/}, {\it Phys. Rev. Lett.\/} {\bf 113}, 133602
	(2014).
	
	\bibitem{Bromley2016}
	S.~L. Bromley, {\it et~al.\/}, {\it Nat. Commun.\/} {\bf 7}, 11039.
	
	\bibitem{Rui2020a}
	J.~Rui, {\it et~al.\/}, {\it Nature\/} {\bf 583}, 369 (2020).
	
	\bibitem{Dicke1954}
	R.~H. Dicke, {\it Phys. Rev.\/} {\bf 93}, 99 (1954).
	
	\bibitem{Purcell1946}
	E.~M. Purcell, {\it Phys. Rev.\/} {\bf 69}, 681 (1946).
	
	\bibitem{Oliver1999}
	W.~D. Oliver, {\it Science\/} {\bf 284}, 299 (1999).
	
	\bibitem{Henny1999}
	M.~Henny, {\it Science\/} {\bf 284}, 296 (1999).
	
	\bibitem{Mueller2010}
	T.~M{\"u}ller, {\it et~al.\/}, {\it Phys. Rev. Lett.\/} {\bf 105}, 040401
	(2010).
	
	\bibitem{Sanner2010}
	C.~Sanner, {\it et~al.\/}, {\it Phys. Rev. Lett.\/} {\bf 105}, 040402 (2010).
	
	\bibitem{Omran2015}
	A.~Omran, {\it et~al.\/}, {\it Phys. Rev. Lett.\/} {\bf 115}, 263001 (2015).
	
	\bibitem{Thomas2016}
	R.~Thomas, {\it et~al.\/}, {\it Nat. Commun.\/} {\bf 7}, 12069 (2016).
	
	\bibitem{Helmerson1990a}
	K.~Helmerson, M.~Xiao, D.~Pritchard, {\it International Quantum Electronics
		Conference\/} (Optical Society of America, 1990), p. QTHH4.
	
	\bibitem{Busch1998}
	T.~Busch, J.~Anglin, J.~Cirac, P.~Zoller, {\it Europhys. Lett.\/} {\bf 44}, 1
	(1998).
	
	\bibitem{Shuve2009}
	B.~Shuve, J.~Thywissen, {\it J. Phys. B: At. Mol. Opt. Phys.\/} {\bf 43},
	015301 (2009).
	
	\bibitem{Morice1995}
	O.~Morice, Y.~Castin, J.~Dalibard, {\it Phys. Rev. A\/} {\bf 51}, 3896 (1995).
	
	\bibitem{Ruostekoski1999}
	J.~Ruostekoski, J.~Javanainen, {\it Phys. Rev. Lett.\/} {\bf 82}, 4741 (1999).
	
	\bibitem{DeMarco1998}
	B.~DeMarco, D.~Jin, {\it Phys. Rev. A\/} {\bf 58}, R4267 (1998).
	
	\bibitem{DeMarco1999}
	B.~DeMarco, D.~S. Jin, {\it Science\/} {\bf 285}, 1703 (1999).
	
	\bibitem{Ruostekoski2009}
	J.~Ruostekoski, C.~J. Foot, A.~B. Deb, {\it Phys. Rev. Lett.\/} {\bf 103},
	170404 (2009).
	
	\bibitem{Bjorklund1980}
	G.~C. Bjorklund, {\it Opt. Lett.\/} {\bf 5}, 15 (1980).
	
	\bibitem{Deb2013}
	A.~B. Deb, B.~J. Sawyer, N.~Kj{\ae}rgaard, {\it Phys. Rev. A\/} {\bf 88},
	063607 (2013).
	
	\bibitem{Deb2020}
	A.~B. Deb, J.~Chung, N.~Kj{\ae}rgaard, {\it New J. Phys.\/} {\bf 22}, 073017
	(2020).
	
	\bibitem{Ketterle}
	W Ketterle, M. W. Zwierlein, “Making, probing and understanding ultracold
	{Fermi} gases” in {\it Ultracold Fermi Gases}, M. Inguscio, W. Ketterle, C.
	Salomon, Eds. (Amsterdam, IOS Press, 2008).
	
\end{thebibliography}

\section*{Author Contributions}

The experimental work and data analysis were done by A. B. D. Both authors contributed to interpreting the results and writing the manuscript based on a first draft written by A. B. D.

\clearpage


\end{document}